%% file: ecocast.tex
\documentclass[10pt,conference,USletter]{IEEEtran}
\ifCLASSOPTIONcompsoc
\usepackage[nocompress]{cite}

\else
\usepackage{cite}
\usepackage{times}
\usepackage{amssymb}
\usepackage{amsmath} 
\usepackage{balance}
\usepackage{amsthm}
\usepackage{comment}
\usepackage{xspace}
\usepackage{subfigure}
\usepackage[ruled]{algorithm}
\usepackage{setspace}%
\usepackage{algorithmic}
\usepackage{graphicx}
\def\ie{\textit{i.e.}\xspace}

\newtheorem{theorem}{Theorem}

\fi
\ifCLASSINFOpdf
\else
\fi
\begin{document}
	%
	\title{ENGINE:Cost Effective Offloading in Mobile Edge Computing with Fog-Cloud Cooperation}

%
\author{\IEEEauthorblockN{Long Chen\IEEEauthorrefmark{1}, Jigang Wu\IEEEauthorrefmark{1},  Xin Long\IEEEauthorrefmark{1}, Zikai Zhang\IEEEauthorrefmark{1} } \IEEEauthorblockA{\IEEEauthorrefmark{1}School of Computer Science and Technology\\
	Guangdong University of Technology,	Guangzhou, Guangdong 510006\\ Email: lonchen@mail.ustc.edu.cn, asjgwucn@outlook.com, longyiyuan@outlook.com, karezhanggdpu@gmail.com}
}
	\maketitle	
\begin{abstract}
\input{abstract}

\end{abstract}

	
	%
	\IEEEpeerreviewmaketitle
	
	\section{Introduction}\label{sec:intro}
	\input{intro.tex}

    \section{Related Work}\label{sec:realted}
    \input{relate.tex}
    \section{System Model and Computational Model}
    \input{sysmodel.tex}\label{sec:sysmodel}
    \section{Problem Formulation}\label{sec:problem}
    \input{prob.tex}
    \section{Analysis for ENGINE Problem}\label{sec:algo}
    \input{algo.tex}

    \section{Algorithms for ENGINE Problem}\label{sec:algoimp}
    \input{algoimp.tex}
    \section{Performance Evaluation}\label{sec:simu}
    \input{simulation.tex}

    \section{Conclusions}\label{sec:conclude}
    \input{conclude.tex}
    	
	
\bibliographystyle{IEEEtran}
\balance
\bibliography{refs}
\end{document}

%% file: abstract.tex
Mobile Edge Computing (MEC) as an emerging paradigm utilizing cloudlet or fog nodes to extend remote cloud computing to the edge of the network, is foreseen as a key technology towards next generation wireless networks. By offloading computation intensive tasks from resource constrained mobile devices to fog nodes or the remote cloud, the energy of mobile devices can be saved and the computation capability can be enhanced. For fog nodes, they can rent the resource rich remote cloud to help them process incoming tasks from mobile devices. In this architecture, the benefit of short computation and computation delay of mobile devices can be fully exploited. However, existing studies mostly assume fog nodes possess unlimited computing capacity, which is not practical, especially when fog nodes are also energy constrained mobile devices. To provide incentive of fog nodes and reduce the computation cost of mobile devices, we provide a cost effective offloading scheme in mobile edge computing with the cooperation between fog nodes and the remote cloud with task dependency constraint. The mobile devices have limited budget and have to determine which task should be computed locally or sent to the fog. To address this issue, we first formulate the offloading problem as a task finish time minimization problem with given budgets of mobile devices, which is NP-hard. We then devise two more algorithms to study the network performance. Simulation results show that the proposed greedy algorithm can achieve the near optimal performance. On average, the Brute Force method and the greedy algorithm outperform the simulated annealing algorithm by about $28.13\%$ on the application finish time.

%% file: intro.tex
Mobile devices such as smartphones, tablets and laptops, are gaining enormous popularity with their capabilities of mobile and portability. As expected, they are playing the leading roles to support various computation intensive applications such as mobile gaming and augmented reality \cite{Grubert2016Towards}. However, such applications are usually delay sensitive and require high computing resources such as power, memory and battery life that frequently exceeds mobile devices can bear. Due to the physical small size, mobile devices are usually constrained by limited computing power \cite{zhao2015energy}, which has become one of the most challenging issues \cite{wang2016phone,gao2017opportunistic,song2014qoi}.

With the growing traffic data into the wireless communication networks such as WiFi, 3G/4G and the emerging 5G, mobile cloud computing (MCC) is designated as a promising solution to address such challenge. By offloading computationally intense tasks to the cloud, which can be viewed as a self managing data center with ample resources, the computing capabilities of mobile devices can be extended \cite{Dinh2013A}. To offload the tasks, the data have to be transmitted from devices to the cloud through wireless communication channels with techniques like network virtualization \cite{Jain2013Network}. A mobile application can be partitioned into multiple sub-tasks with task dependency. The sub-tasks can be executed either on the mobile device its-self locally or onto the remote cloud. With this setting, by carefully select tasks for remote execution, the lifetime of mobile devices can be prolonged and user experiences can be enhanced.

Although MCC enables convenient access  to a pool of computation resources in the cloud, moving all the tasks on mobile devices to the remote cloud would result in large transmission latencies that degrade the Quality of Experience (QoE) of mobile users. Mobile Edge Computing (MEC) or fog computing \cite{hu2015mobile} has recently emerged as a remedy to the above limitations. By deploying fog or cloudlet nodes that are closer to mobile users at the edge of the network, mobile users can share the same services as the remote cloud. Whereas the transmission delay can be reduced while meeting the computation resource demands of mobile devices with MEC. For example, in the heterogeneous wireless network, small cell base stations can be deployed with fog nodes to serve local mobile users \cite{Felemban2013A}. The fog nodes can be any devices with storage, computing capabilities and network interfaces. In a local community, fog nodes can be deployed at shopping centers, hotels or even bus-stops with WiFi access and deliver computing results back to their mobile users. Although fog computing demonstrates its potential to improve the QoE of mobile users by bring services close to users, fog nodes themselves can be resource constrained. When burst traffic arrives, fog nodes on their own may not be able to serve users. Therefore, the remote cloud resources can be borrowed by fog nodes via fog-cloud cooperation.

It should be noted that MEC with fog-cloud cooperation promises enormous benefits, designing energy efficient schemes for computation offloading should answer the following questions. (i) Which sub task should be executed locally on the mobile device and which should be offloaded? (ii) How much moneytary compensation should be paid by mobile users to stimulate the offloading of fog nodes and remote cloud? (iii) Which tasks should be migrated to which remote cloud server by the fog such that the total cost is within the threshold of mobile devices?

To answer the above questions, in this paper, we concentrate on the cost Effective offloadiNg in mobile edGe computINg with fog-cloud coopEration (ENGINE) problem, in which the following issues will be addressed. Firstly, for an application with multiple sub-tasks that follow task-dependency, the offloading strategy adopted by the precedence task can affect the successor's action. Secondly, the remote cloud is abundant in storage and computation resources with long delay while the fog nodes are resource constrained with short latencies. Therefore, the coordination between fog and remote cloud servers should be carefully designed to meet the QoE demands of mobile users. Thirdly, to guarantee the QoE of mobile users, the cost constraint of mobile devices should be taken into account in designing offloading strategies.

In this paper, the objective is to design a cost effective computation offloading and resource scheduling scheme with given cost budget of the mobile device running an application. Compared to existing work \cite{farris2017lightweight,Chen2017Dynamic,lizhen2015genetic,Kao2015Hermes,mahmoodi2016optimal}, the main contribution of this paper is summarized as follows.

\begin{itemize}
	\item Taking task dependency into consideration, the detailed task execution procedures, such as which task should be executed on the mobile device and which task should be offloaded, how to determine which task to be further offloaded to which remote cloud server.
	\item The ENGINE problem is formulated as a response time minimization problem under the constraints of cost budgets and task-precedence requirements.
	\item To solve the optimization problem, we propose a distributed algorithm for the joint optimal offloading task selection and fog-cloud cooperation. Extensive experiments demonstrate the effectiveness of proposed schemes.
	\item  To the best of authors' knowledge, it is the first to consider the joint device-fog-cloud scheduling and offloading work that minimizing the execution delay of the application.
\end{itemize}

The remainder paper is organized as follows. Related works on offloading in MEC are presented in Section \ref{sec:realted}. Section \ref{sec:sysmodel} presents the system model and computational model. The problem formulation of ENGINE problem is described in Section \ref{sec:problem}. Section \ref{sec:algo} presents the distributed algorithm for ENGINE. The performance evaluation is presented in Section \ref{sec:simu} and Section \ref{sec:conclude} concludes this paper followed by the future work.

%% file: relate.tex
Computation offloading in MCC has been extensively studied in the literature with a variety of architectures and offloading policies \cite{Chen2017Dynamic} \cite{Kosta2012ThinkAir} \cite{ sanaei2014heterogeneity}. However, implementing such offloading invokes extra communication delay due to the long distance of remote cloud servers. Instead of conventional MCC, MEC as defined by the European Telecommunications Standards Institute (ETSI) \cite{hu2015mobile} is widely recognized as a key technology for the next generation network.

In MEC, computation offloading can be basically classified into two types, \textit{i.e.} full offloading \cite{Kamoun2015Joint,Chen2016Efficient,souza2016handling} and partial offloading \cite{you2016multiuser,wang2016mobile,guo2016index,farris2017optimizing}. For full offloading, the whole computation tasks are offloaded and processed by the MEC. In \cite{Kamoun2015Joint}, based on Markov decision process, Kamoun et al. proposed both online learning and off-line schemes to minimize energy consumption of mobile devices by offloading all packets to edge cellular base stations with delay constraints. Chen et al. in \cite{Chen2016Efficient} proposed a game theoretic offloading scheme in the multi-channel wireless contention environment. Souza et al. in \cite{souza2016handling} studied the service allocation problem with the objective of minimizing total delay of resources allocation. For partial offloading, part of the computation tasks are processed locally on the mobile devices while the rest are offloaded to the MEC. In \cite{you2016multiuser}, by using convex optimization, You et al. presented multi-user offloading algorithms to reduce the energy consumption of mobile devices with delay constraint. In \cite{wang2016mobile}, Wang et al. provided a dynamic voltage scaling based partial offloading scheme. Similarly as \cite{Kamoun2015Joint}, Guo et al. \cite{guo2016index} presented a discrete-time Markov decision process to achieve optimal power-delay tradeoff. Moreover, Farris et al.  \cite{farris2017optimizing} \cite{farris2017lightweight} proposed QoE guaranteed service replication for delay sensitive applications in 5G edge network. However, they all focus on how much workload should be distributed to the MEC without considering task dependency for an application.

Recently, there have been some works on computation offloading with task dependency in MEC \cite{tziritas2017data,rimal2017workflow}. In \cite{tziritas2017data}, Tziritas et al. proposed a data replication based virtual machine migration scheme in edge network. In \cite{rimal2017workflow}, under the multi-tenant cloud computing environment, Rimal et al. designed a few algorithms to schedule workflows with flow delay constraints. However, both of \cite{tziritas2017data} and \cite{rimal2017workflow} are not suitable for the scenario which is investigated by \cite{mahmoodi2016optimal} in MCC where applications with sub-tasks that must be executed on the mobile device. In this paper, for an application, we investigate partial offloading scheme with joint consideration of computation cost of mobile devices and the fog nodes.

To the best of authors' knowledge, there is only a little work that has addressed the computation offloading problem in MEC taking into account the cost of edge servers such as energy consumption and related communication. In \cite{deng2015towards}, Deng et al. investigated the power consumption and delay tradeoff with the objective to minimize total system power consumption of fog nodes and remote cloud servers. Compared to our work, however, they failed to consider the task dependency and cost budgets of mobile devices. In this paper, we conduct offloading study from the perspective of mobile users. 

%% file: sysmodel.tex
This section firstly describes the system model and formulates the ENGINE problem with local computing, fog computing and fog-cloud cooperation.

\subsection{System Model}
\begin{figure}[h]
\includegraphics[width=2.5in]{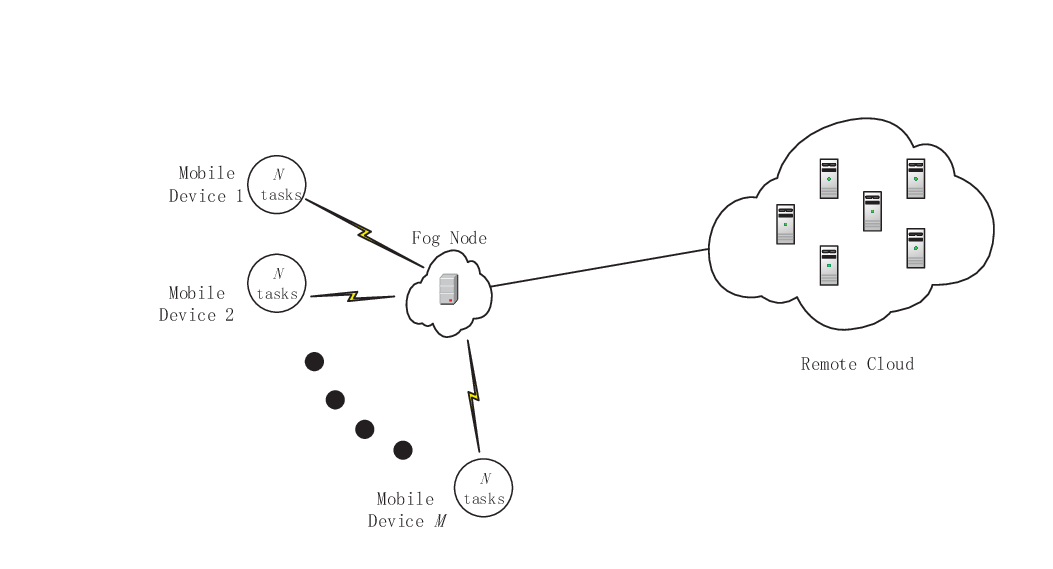}
\caption{System Architecture}\label{fig:network}
\end{figure}
As shown in Fig. \ref{fig:network}, we assume a group of mobile devices denoted as $\mathcal{M}=\{1,2,\cdots,M\}$ which are located in the vicinity of its corresponding wireless access points. Each access point can be a WiFi or a small cell base station in a HetNet and fog nodes are connected with those access points. The access points are connected with each other via wired links with wich the remote cloud servers are connected. A mobile application in MEC is partitioned into $N$ sub-tasks, denoted by a set of $\mathcal{N}=\{1,2,\cdots,N\}$.  Let $\mathcal{K}_{c}=\{1,2,\cdots,P\}$ and $\mathcal{K}_{f}=\{1,2,\cdots,Q\}$ denotes the set of cloud servers and the set of fog servers respectively. There is a set of $\mathcal{K}=\mathcal{K}_{c}\cup\mathcal{K}_{f}$ cloud nodes in total since we assume there are more remote servers than fog nodes. Hence, $|\mathcal{K}|=P+Q$.

The application is modeled as a weighted Directed Acyclic Graph (DAG) $G(V,E)$, where the set $V$ of vertices $V=\{v_i|i=1,2,\cdots,N\}$ denotes sub-tasks of the application with $|V|=N$ and $E=\{e_{ij}|(i,j)\in \{1,2,\cdots,N\}\times\{1,2,\cdots,N\}\}$, $|E|=e$ is the set of communication edges representing the precedence relation such that task $v_i$ should complete its execution before task $v_j$ starts.

Next, we introduce the communication and computation models for mobile devices, fog nodes and between fog nodes and remote cloud servers in detail.

\subsection{Wireless Communication Model}
We first present the wireless communication model and then provide the wired communication model.

The channel from mobile device $ m \in \mathcal{M} $ to access point $p \in
\mathcal{K}_f$ follows quasi-static block fading. We let $x_{m,n,p}^{f}\in\{0,1\}$, $k\in \mathcal{K}_{f}$ and $x_{m,n,q}^{c}\in\{0,1\}$, $q\in \mathcal{K}_{c}$ denote the computation offloading strategy made by the mobile device. Particularly, $x_{m,n,p}^{f}=1$ means that mobile device $m$ chooses to offload the computation task $n$ to the $p$th fog node in MEC while $x_{m,n,p}^{f}=0$ implies that device $m$ chooses to execute the $n$th task locally on its own device. Similarly, $x_{m,n,q}^{c}=1$ means the subtask $n$ is further offloaded to the $q$th remote cloud server and $x_{m,n,q}^{c}=0$ does not. We can compute the uplink data rate for wireless communication with access point $p$, $p\in\mathcal{K}_{f}$ as
\begin{equation}\label{eq:CalRate}
R_{m,n,p}=  W\log_2\left(1+\frac{P_{m,n,p}^{tx}h_{m,n,p}}{\sigma_{m,p}^{2}+\sum\limits_{i\neq m,j\neq n}P_{i,j,p}^{tx}h_{i,j,p}}\right),
\end{equation}
where $x_{i,j,p}^{f}=1$, and $P_{m,n,p}^{tx}$ is the transmission power of mobile device $m$ to offload its task $n$ to the access point $p$. The channel gain from the $m$th device to the $p$th access point is $h_{m,n,p}$ when transmitting task $n$. The channel bandwidth is $W$ and the surrounding noise power at the receiver with the transmission link $(m,p)$ is $\sigma_{m,p}^{2}$. From \ref{eq:CalRate}, we can see the transmission rate is in proportion to the transmission power of mobile devices and is in inverse proportion to the interference power of neighbouring devices.

\subsection{Computation Model}
 Let $\tau_{m,n}^{l}$ be the completion time of local execution of task $n$ on device $m$. Let $\tau_{m,n}^{t}$ be the transmission time between the mobile device and its corresponding access point, and $\tau_{m,n}^{f}$ be the execution time on fog node. Denote $\tau_{m,n,q}^{r}$ as the transmission time between the fog nodes and the remote cloud server $q$. Denote $\tau_{m,n,q}^{c}$ as the execution time of task $n$ of device $m$ on remote cloud server $q$. Next, we present the computation overhead on energy consumption, task completion time as well as the coordination between fog nodes and remote cloud.

 \subsubsection{Local Computing}
 Let $f_m^l$ be the computation capability, \textit{i.e.} the CPU clock speed (cycles/second) of mobile devices. We note different mobile devices may have different CPU clock speed. The computation execution time of task $n$ on mobile device $m$ is then calculated as
 \begin{equation}\label{eq:taumnlocal}
\tau_{m,n}^{l}=\frac{w_{m,n}}{f_m^l},
 \end{equation}
 and the energy consumption of mobile device $m$ for the corresponding task $n$ is given by
 \begin{equation}\label{eq:enermnlocal}
 E_{m,n}^{l}={\kappa}w_{m,n}{f_{m}^{l}}^2,
 \end{equation}
 where $\kappa$ is set to be $10^{-11}$ according to \cite{guo2016energy} and $w_{m,n}$ is the workload of task $n$ of mobile device $m$.
\subsubsection{Fog and Cloud Computing}
Similar to \cite{guo2016energy}, we ingnore the time and energy consumption that the cloud returns the computation outcome back to the user. That's because for many applications, the size of the outcome usually small. Let $f_{p}^{f}$ be the computation ability of fog node $p$, \textit{i.e.} the machine CPU frequency of the fog node. Then the computation execution time is given by
\begin{equation}
\tau_{m,n}^{f}=\frac{w_{m,n}}{f_p^f}
\end{equation}
and the energy consumption of the fog node is given by
\begin{equation}\label{eq:enermnfog}
E_{m,n}^{f}=\left(\alpha_f \left({f_p^{f}}\right)^{\epsilon}+\beta_f \right) \tau_{m,n}^{f},
\end{equation}
where $\alpha_f$ and $\beta_f$ are positive constants which can be obtained by offline power fitting and the value $\epsilon$ ranges from $2.5$ to $3$ \cite{rao2012distributed}. Similarly, the energy consumption of the remote cloud server for the task $n$ of mobile device $m$ is given by
\begin{equation}\label{eq:enermncloud}
E_{m,n}^{c}=\left(\alpha_c \left(f_q\right)^{\epsilon}+\beta_c\right)\tau_{m,n,q}^{c},
\end{equation}
where the remote execution time $\tau_{m,n,q}^{c}$ is calculated as
\begin{equation}
\tau_{m,n,q}^{c}=\frac{ w_{m,n}}{f_q^{c}}.
\end{equation}
Similarly, $f_q^c$ denotes the CPU frequency of the $q$th remote cloud server and $\alpha_c$, $\beta_c$ are also positive constants.

\subsubsection{Data Transferring Cost} Given the $n$th task size of mobile device $m$, including the input data \textit{i.e.} $d_{m,n}$, $\tau_{m,n}^{t}$ can be expressed as
\begin{equation}\label{eq:taumnt}
\tau_{m,n}^{t}= \frac{d_{m,n}}{R_{m,n,p}}.
\end{equation}
Then the energy cost when transferring the data to the access point is given by
\begin{equation}\label{eq:enertrantoap}
E_{m,n}^{t}=P_{m,n,p}^{tx} \tau_{m,n}^{t}.
\end{equation}
Further more, we can get the delay data transfer if the remote server is employed by the corresponding fog node as
\begin{equation}\label{eq:delayfogcloud}
\tau_{m,n,q}^{r}=\frac{d_{m,n}}{\omega},
\end{equation}
where $\omega$ is the average transfer rate or bandwidth between fog node and the corresponding remote server. The energy cost of fog node during the offloading to the remote server which is denoted by $E_{m,n}^{s}$, is given by
\begin{equation}\label{eq:transfercostfog}
E_{m,n}^{s}=P_0 \cdot \tau_{m,n,q}^{r},
\end{equation}
where $P_0$ is the amount of additional power when performing data transfer per unit time from fog node to the remote server.

\subsubsection{Basic Contraints}
Before we formulate the ENGINE problem, we present some definitions, QoS as well as user budget constraints.  Note that for a particular task of a mobile user, it cannot be executed unless all its precedence tasks have already been processed. We name this constraint as precedence constraint followed by \cite{Chen2017Dynamic} and \cite{mahmoodi2016optimal}. Let $TR_{m,n}^{l}$ be the time when task $n$ of mobile device $m$ is ready to be processed. Then we have
\begin{equation}\label{eq:TRlmneq}
TR_{m,n}^{l}=  \max_{ k\in pre(n)} \max \{ TF_{m,k}^{l},  TF_{m,k}^{f},   TF_{m,k}^{c} \},
\end{equation}
where the receiving delay is neglected following \cite{guo2016energy} and (\ref{eq:TRlmneq}) can be rewritten as
\begin{equation}\label{eq:trmnl}
\begin{split}
 TR_{m,n}^{l}\ge  &  (1- x_{m,k,q}^{c})\left[ \left(1-x_{m,k,p}^{f}\right) {TF}_{m,k}^{l}+x_{m,k,p}^{f}TF_{m,k}^{f}\right]  \\
 &  + x_{m,k,q}^{c} TF_{m,k}^{c} ,    \quad \quad {k\in pre(n)},
 \end{split}
\end{equation}
where $pre{(n)}$ denotes the prececessors of task $n$, $k\in pre(n)$ in (\ref{eq:trmnl}) means the local computing of task $n$ can be executed only after task $k$ has been executed. Therefore, the local task completion time of mobile device $m$, which is denoted by $TF_{m,n}^{l}$ is given by
\begin{equation}\label{eq:tfmnl}
TF_{m,n}^{l}= \tau_{m,n}^{l}+ TR_{m,n}^{l}.
\end{equation}
Similarly, let $TR_{m,n}^{f}$, $TR_{m,n}^{c}$ be the time when task $n$ of mobile device $m$ is ready to be processed on the fog node and the corresponding remote cloud server, given by
\begin{equation}\label{eq:trmnf}
TR_{m,n}^{f}=\max\{TF_{m,n}^{t},  \max_{k\in pre(n)} TF_{m,k}^{f}, \max_{k\in pre(n)} TF_{m,k}^{c} \},
\end{equation}
and
\begin{equation}\label{eq:trmnc}
\begin{split}
 TR_{m,n}^{c} =  & \max \left\{TF_{m,n}^{t}+\tau_{m,n}^{r}, \max \limits_{k\in pre(n)} TF_{m,k}^{c} ,  TF_{m,n}^{r}    \right\},
  \end{split}
\end{equation}
In (\ref{eq:trmnf}), ${TF}_{m,n}^{t}$ is the finished transmission time from mobile device $m$ to the corresponding fog node and is given by
\begin{equation}\label{eq:tfmnt}
TF_{m,n}^{t}=\tau_{m,n}^{t}+ \max_{ k\in pre(n)}TF_{m,k}^{l},
\end{equation}
$TF_{m,n}^{f}$ is the task finish time on the fog node, given by
\begin{equation}\label{eq:tfmnf}
TF_{m,n}^{f}= \tau_{m,n}^{f}+TR_{m,n}^{f},
\end{equation}
and $TF_{m,k}^{c}$ is the task finish time on remote cloud server, given by
\begin{equation}\label{eq:tfmnc}
TF_{m,n}^{c}=\tau_{m,n,q}^{c}+TR_{m,n}^{c}.
\end{equation}

In (\ref{eq:trmnc}), $TF_{m,n}^{r}$ is the completion time of transmission between fog and remote cloud server, which can be defined as similar as (\ref{eq:tfmnt}), given by
\begin{equation}\label{eq:TFmnr}
TF_{m,n}^{r}=\tau_{m,n}^{r}+\max_{ k\in pre(n)}TF_{m,k}^{f}.
\end{equation}
In (\ref{eq:tfmnc}), $TR_{m,n}^{c}$ is the time when the task $n$ is ready for processing at the remote cloud server.
	
It can be observed from (\ref{eq:trmnf}) that if the predecessor task $k$ of $m$ is executed locally, then $TF_{m,k}^{f} =0$, $TF_{m,k}^{c}=0$, $TF_{m,n}^{t}=0$. The term $\max \limits_{k\in pre(n)} TF_{m,k}^{f} $ indicates all the predecessors of task $n$ that are offloaded to the fog node have finished execution and $\max \limits_{k\in pre(n)}   TF_{m,k}^{c}$ means all the predecessors of task $n$ that on the remote cloud server have finished execution as well. Therefore, the precedence constraints in (\ref{eq:trmnf}) and (\ref{eq:trmnc}) can be rewritten as
\begin{equation}\label{eq:trmnfge}
\begin{split}
& TR_{m,n}^{f} \ge TF_{m,n}^{t},  \\
& TR_{m,n}^{f} \ge \max_{k\in pre(n)}TF_{m,k}^{f}, \\
& TR_{m,n}^{f} \ge \max_{k\in pre(n)} TF_{m,k}^{c},
\end{split}
\end{equation}
and
\begin{equation}\label{eq:trmncge}
\begin{split}
& TR_{m,n}^{c} \ge TF_{m,n}^{t}+\tau_{m,n}^{r} ,\\
& TR_{m,n}^{c} \ge  TF_{m,n}^{r}, \\
& TR_{m,n}^{c} \ge \max \limits_{k\in pre(n)} TF_{m,k}^{c}.
\end{split}
\end{equation}

Next, we derive the utility constraints of fog nodes as well as remote cloud servers  and cost constraints of mobile devices. The utility of fog node can be derived as
\begin{equation}\label{eq:utilityfog}
U_{p}^{f}=\sum_{m=1}^{M} \sum_{n=1}^{N} \left( P_{p}^{f} d_{m,n}x_{m,n,p}^{f}- E_{m,n}^{s}x_{m,n,q}^{c} -E_{m,n}^{f}x_{m,n,p}^{f}\right),
\end{equation}
where $ \forall q\in \mathcal{K}_{c}$ and $P_{p}^{f}$ is the charging price by the $p$th fog node to cover the transmission or execution cost per unit data in the network. It can be observed from (\ref{eq:utilityfog}) that the transmission cost $E_{m,n}^{s}$ and execution cost $E_{m,n}^{f}$ cannot coexist for a particular task $n\in N$. Similarly, the utility of the remote cloud server can be expressed as
\begin{equation}\label{eq:utilitycloud}
U_{q}^{c} = \sum_{m=1}^{M}\sum_{n=1}^{N}\left(P_q^{c}d_{m,n}x_{m,n,q}^{c}-E_{m,n}^{c}x_{m,n,q}^{c}\right), \forall q\in \mathcal{K}_{c}
\end{equation}
where $P_{q}^{c}$ is the charging price of the $q$th remote cloud server. It should be noted that, to motivate computation offloading, the utility of both fog nodes and the remote cloud server should not be negative, therefore we have
\begin{equation}\label{eq:upfqc}
U_{p}^{f} \ge 0, U_{q}^{c} \ge  0, \forall p\in \mathcal{K}_{f}, \forall q\in \mathcal{K}_{c}.
\end{equation}
For the mobile device $m$ processing task $n$, its cost $C_{m,n}$ is consists of local execution cost, the payment for the corresponding fog node and the payment for the remote cloud server, which can be expressed as
\begin{equation}\label{eq:mdcost}
\begin{split}
C_{m,n} =  &  \left\{\left(1-x_{m,n,q}^{c}\right) \left[\right.E_{m,n}^{l}\left(1-x_{m,n,p}^{f}\right)+\right. \\
& \left. P_{p}^{f}d_{m,n}x_{m,n,p}^{f}\left.\right]+ P_{q}^{c}d_{m,n}x_{m,n,q}^{c}   \right\}, \forall p\in \mathcal{K}_f, \forall q\in \mathcal{K}_c,
\end{split}
\end{equation}
where for a particular task $n \in N$ of mobile device $m$, $x_{m,n,q}^{c}$, $x_{m,n,p}^{f}$ cannot be one at the same time.

Finally, we derive the runtime expression of the whole application for the mobile device. Denote $T_m$ as the total application response time for mobile device $m$, then $T_m$ is the time when all the tasks in an application are finished, given by
\begin{equation}\label{eq:Totaldelay}
\begin{split}
T_m=& \sum_{n=1}^{N}\left[ (1-x_{m,n,q}^{c})(1-x_{m,n,p}^{f})TF_{m,n}^{l}+ \right. \\  & \left. (1-x_{m,n,q}^{c})x_{m,n,p}^{f}TF_{m,n}^{f}+ x_{m,n,q}^{c}TF_{m,n}^{c}\right], \forall m\in \mathcal{M}
\end{split}
\end{equation}

We can observe from (\ref{eq:Totaldelay}) that the total application delay is the time when the final task $N$ of mobile device $m$ has been finished on the mobile device.

%% file: prob.tex
In this section, we will formulate the ENGINE problem.

To solve ENGINE problem, taking mobile device budget and the cost of cloud into consideration, we try to design effective computation offloading strategy with fog and remote cloud cooperation. The aim is to minimize the total application response delay. Therefore, according to constraints (\ref{eq:trmnl}), (\ref{eq:trmnfge}), (\ref{eq:trmncge}) and (\ref{eq:upfqc}), the ENGINE problem for all mobile devices can be formulated as a constrained minimization problem as follows
\begin{equation}\label{eq:obj}
\mathbf{OPT-1}\ \ obj: \min \sum_{m=1}^{M}  T_m
\end{equation}
subject to $\forall m\in \mathcal{M}$, $\forall n\in \mathcal{N}$, $\forall p\in \mathcal{K}_f$, $\forall q\in \mathcal{K}_c$,
\begin{equation*}
\begin{split}
C1: \quad & TR_{m,n}^{l}\ge (1- x_{m,k,q}^{c})\left[ \left(1-x_{m,k,p}^{f}\right) {TF}_{m,k}^{l} \right. \\
&\left.+x_{m,k,p}^{f}TF_{m,k}^{f}\right] + x_{m,k,q}^{c} TF_{m,k}^{c} , {k\in pre(n)},
\end{split}
\end{equation*}
\begin{equation*}
\begin{split}
C2:\quad & TR_{m,n}^{f} \ge TF_{m,n}^{t}, \\
&  TR_{m,n}^{f} \ge \max_{k\in pre(n)}TF_{m,k}^{f}, \\
& TR_{m,n}^{f} \ge \max_{k\in pre(n)} TF_{m,k}^{c},
\end{split}
\end{equation*}
\begin{equation*}
\begin{split}
C3:\quad & TR_{m,n}^{c} \ge TF_{m,n}^{t}+\tau_{m,n}^{r}\\
& TR_{m,n}^{c} \ge  TF_{m,n}^{r}, \\
& TR_{m,n}^{c} \ge \max \limits_{k\in pre(n)} TF_{m,k}^{c}, \\
\end{split}
\end{equation*}
\begin{equation*}
\begin{split}
C4:\quad & U_{p}^{f} \ge 0, U_{q}^{c} \ge  0,
\end{split}
\end{equation*}
\begin{equation*}
\begin{split}
C5:\quad & x_{m,n,q}^{c}+x_{m,n,p}^{f}\le 1 ,
\end{split}
\end{equation*}
\begin{equation*}
\begin{split}
C6:\quad & x_{m,n,q}^{c}\in \{0,1\},  x_{m,n,p}^{f} \in \{0,1\},
\end{split}
\end{equation*}
\begin{equation*}
\begin{split}
C7:\quad &\sum_{n=1}^{N} C_{m,n} \le  \epsilon_m,
\end{split}
\end{equation*}

where $\epsilon_m$ is the sum budget for mobile device $m$. Constraint $C1$ is the local task dependency constraint that ensures task $n$ can start to execute only after all its predecessor tasks have finished. Constraints $C2$ and $C3$ are fog and remote cloud task dependency constraints which implies that task $n$ can be executed on the fog and remote cloud server only after the task has been completely offloaded to the fog and remote cloud accordingly. Constraint $C4$ are the utility constraints for fog nodes and the remote cloud server. Constraint $C5$ ensures for a task $n\in \mathcal{N}$, it can only be executed on one of the three places, \textit{i.e.} the local mobile device, the fog node and the remote cloud server. The binary constraints are presented in $C6$ and $C7$ is the budget constraint for mobile device $m$.

%% file: algo.tex
In this section, we solve $\mathbf{OPT-1}$ with the relaxation of constraint $C6$ and give the analysis on ENGINE problem.
\subsection{Lagrangian Relaxation}
The challenge to solve $\mathbf{OPT-1}$ is due to constraint $C6$. That's because the binary constraints in $C6$ make the $\mathbf{OPT-1}$ problem a mixed integer programming problem which is non-convex and NP-hard. Therefore, as mentioned by \cite{boyd2004convex}, we first relax the binary constraints in $C6$ to a real number \textit{i.e.} $0\le x_{m,n,q}^{c}\le 1$ and $0 \le x_{m,n,p}^{f}\le 1$.  Obviously, the $\mathbf{OPT-1}$ problem with the relaxed optimization values $x_{m,n,p}^{f}$ and $x_{m,n,q}^{c}$ is still not convex because the constraint C1 is not convex. Therefore, to solve the problem, we have to solve the Lagrange dual problem of $\mathbf{OPT-1}$, which is usually convex and provides a lower bound of $\mathbf{OPT-1}$ on the optimal value \cite{boyd2004convex}. By relaxing the task ready time constraints with non-negative Lagrangian mulipliers, we first derive the Lagrangian Relaxed (LR) function of the primal problem $\mathbf{OPT-1}$. It is worthy noting that the term $x_{m,k,q}^{c}$ and $x_{m,k,p}^{f}$ makes the constraint C1 and the object function (\ref{eq:obj}) nonconvex, to deal with the issue, we introduce a variable $x_{m,n}^{a}=x_{m,n,p}^{f}\cdot x_{m,n,q}^{c}$, $0\le x_{m,n}^{a}\le 1$, the Lagrangian relaxed function of (\ref{eq:obj}) is formulated as (\ref{eq:LR}).
\begin{equation}\label{eq:LR}
\centering
\begin{split}
&(LR)  L(x_{m,n,p}^{f},x_{m,n,q}^{c},x_{m,n}^{a},\mu,\omega,\lambda,\psi,\phi,\gamma) =\min \sum_{m=1}^{M} T_{m_L } \\
=& \sum_{m=1}^{M} \sum_{n=1}^{N}\left[ (1-x_{m,n,q}^{c}-x_{m,n,p}^{f}+  x_{m,n}^{a})TF_{m,n}^{l}+\right. \\
 & (x_{m,n,p}^{f}-x_{m,n}^{a}) \left. TF_{m,n}^{f}+x_{m,n,q}^{c}TF_{m,n}^{c}\right]+ \\
& \mu_{m,n}(TF_{m,n}^{t}-TR_{m,n}^{f})+\omega_{m,n}(TF_{m,n}^{t}+\tau_{m,n}^{r}-TR_{m,n}^{c})+ \\
&\lambda_{m,n}(TF_{m,n}^{r}-TR_{m,n}^{c}) - \psi_{m,n} U_{p}^{f}-\phi_{m,n} U_{q}^{c}+ \\
&\gamma_{m,n}\left(\sum_{n=1}^{N}C_{m,n}-\epsilon_m\right),
\end{split}
\end{equation}
\begin{equation*}
\begin{split}
\textit{s.t.} & \quad \quad \quad  0\le x_{m,n,q}^{c}+x_{m,n,p}^{f}\le 1, \quad \quad  (C11) \\
& \quad \quad \quad \quad 0\le x_{m,n,p}^{f} \le 1,  \quad \quad \quad \quad \quad (C12)\\
& \quad \quad \quad \quad 0\le x_{m,n,q}^{c} \le 1 , \quad \quad \quad \quad \quad (C13)\\
& \quad \quad \quad \quad 0 \le x_{m,n}^{a} \le x_{m,n,p}^{f},  \quad \     \quad \quad (C14) \\
&  \quad \quad \quad \quad  0\le x_{m,n}^{a} \le x_{m,n,q}^{c},  \quad  \  \quad \quad(C15)\\
\end{split}
\end{equation*}
\begin{equation*}
\begin{split}
&  \quad \quad \quad TR_{m,n}^{f} \ge \max_{k\in pre(n)}TF_{m,k}^{f}, \quad  \quad (C16) \\
& \quad \quad \quad  TR_{m,n}^{f} \ge \max_{k\in pre(n)} TF_{m,k}^{c}, \quad  \quad (C17) \\
& \quad \quad \quad TR_{m,n}^{c} \ge \max \limits_{k\in pre(n)} TF_{m,k}^{c},  \quad \quad (C18)\\
\end{split}
\end{equation*}

It should be noted that constraints C14 and C15 in (\ref{eq:LR}) is relaxed from $x_{m,n}^{a}=x_{m,n,p}^{f}\cdot x_{m,n,q}^{c}$ and $0\le x_{m,n}^{a} \le 1$ to ensure the convexity, and
\begin{equation}\label{eq:LM}
\mu_{m,n}, \omega_{m,n}, \lambda_{m,n}, \psi_{m,n},\phi_{m,n},\gamma_{m,n} \ge 0,
\end{equation}
\subsection{Dual Problem Formulation}
The corresponding Lagrangian Dual (LD) problem is formulated as follows:
\begin{equation}\label{eq:LD}
\begin{split}
(LD)\quad \max \ L(\mu,\omega,\lambda,\psi,\phi,\gamma)=\max \  \min T_{m_L},
\end{split}
\end{equation}
subject to constraints C11-C15 and (\ref{eq:LM}). The dual problem is decomposed into two layers \textit{i.e.}, the inner layer minimization in (\ref{eq:LD}) with $M$ subproblems that can be executed in parallel since there are $M$ mobile users and the outer layer maximization problem. In the following, we give the distributed solution of the computation offloading selection and transmission power allocation.
\subsection{Computation Offloading Decision}
In the computation offloading decision procedure, the system will determine which subtask should be executed on the mobile device, which subtask should be offloaded to the fog node and which further be transmitted to the remote cloud server by the fog node. The objective is to determine the strategy to minimize execution delay with given budget constraints. Meanwhile, the task dependency cosntraints should be preserved. The optimal computation offloading decision subproblem can be obtained by solving the following minimization problem:
\begin{equation}\label{eq:cod}
\begin{split}
&\min_{x_{m,n,p}^{f},x_{m,n,q}^{c},x_{m,n}^{a}} \sum_{n=1}^{N}\left[(1-x_{m,n,q}^{c}-x_{m,n,p}^{f}+x_{m,n}^{a})TF_{m,n}^{l}+ \right. \\
& \left. (x_{m,n,p}^{f}-x_{m,n}^{a})TF_{m,n}^{f}+x_{m,n,q}^{c}TF_{m,n}^{c}\right]+\\
& \mu_{m,n}(TF_{m,n}^{t}-TR_{m,n}^{f})+\omega_{m,n}(TF_{m,n}^{t}+\tau_{m,n}^{r}-TR_{m,n}^{c})+\\
& \lambda_{m,n}(TF_{m,n}^{r}-TR_{m,n}^{c}), \\
\end{split}
\end{equation}
subject to constraints C2\--C4, C7 and C11\--C18.

It can be observed from (\ref{eq:cod}) that if $TF_{m,n}^{l}<TF_{m,n}^{f}$ and $TF_{m,n}^{l}<TF_{m,n}^{c}$, the subtask $n$ of mobile user $m$ will be executed locally on the mobile phone. Then object function will achieve the minimum value if $x_{m,n,q}^{c}=0$, $x_{m,n,p}^{f}=0$. If $TF_{m,n}^{f}<TF_{m,n}^{l}$ and $TF_{m,n}^{f}<TF_{m,n}^{c}$, then the task will be offloaded to the fog node. If $TF_{m,n}^{c}<TF_{m,n}^{l}$ and $TF_{m,n}^{c}<TF_{m,n}^{f}$, the remote cloud will be chosen. Note that $\mu_{m,n}$, $\omega_{m,n}$, $\lambda_{m,n}$ will we zero when the minium value of (\ref{eq:cod}) is obtained.

\subsection{Optimal Power}
The optimal power allocation strategy tries to allocate the power for mobile devices to minimize the total task completion delay with budget constraints. It is obviously that the strategy is valid when $x_{m,n,p}^{f}=1$ and $x_{m,n,q}^{c}=0$ or $x_{m,n,p}^{f}=0$ and $x_{m,n,q}^{c}=1$. Therefore, the power allocation scheme can be obtained by solving the following minimization subproblem:
\begin{equation}\label{eq:opa1}
\min_{P_{m,n,p}^{tx}}\sum_{m=1}^{M} \sum_{n=1}^{N}  \left\{TF_{m,n}^{f}+\mu_{m,n}(TF_{m,n}^{t}-TR_{m,n}^{f})\right\},
\end{equation}
subject to constraints  C2\--C4, C7 and C11\--C18, when fog computing is adopted, \textit{i.e.} $x_{m,n,p}^{f}=1$ and $x_{m,n,q}^{c}=0$. When remote cloud computing happens, \textit{i.e.} $x_{m,n,p}^{f}=0$ and $x_{m,n,q}^{c}=1$, the objective becomes
\begin{equation}\label{eq:opa2}
\begin{split}
 \min_{P_{m,n,p}^{tx}} &\sum_{m=1}^{M} \sum_{n=1}^{N}  \left\{TF_{m,n}^{c}+\omega_{m,n}(TF_{m,n}^{t}+ \right.\\
& \left. \tau_{m,n}^{r}-TR_{m,n}^{c})+\lambda_{m,n}(TF_{m,n}^{r}-TR_{m,n}^{c})\right\},
\end{split}
\end{equation}
subject to constraints  C2\--C4, C7 and C11\--C18. For (\ref{eq:opa1}), there are three cases.

In (\ref{eq:opa1}), there are three cases listed as follows.

\textbf{Case I:} If $TF_{m,n}^{t}>\max_{ k\in pre(n)}TF_{m,k}^{f}$ and $TF_{m,n}^{t}>\max_{ k\in pre(n)}TF_{m,k}^{c}$, we have $TR_{m,n}^{f}=TF_{m,n}^{t}$. Therefore, the objective function can be rewritten as:
\begin{equation}\label{eq:case1}
F(P_{m,n,p}^{tx})= \sum_{m=1}^{M} \sum_{n=1}^{N} TF_{m,n}^{f} = \sum_{m=1}^{M} \sum_{n=1}^{N} (\tau_{m,n}^{f}+TF_{m,n}^{t}).
\end{equation}

\textbf{Case II:} If $\max_{ k\in pre(n)}TF_{m,k}^{f}>TF_{m,n}^{t}$ and $\max_{k\in pre(n)}TF_{m,k}^{f}>\max_{ k\in pre(n)}TF_{m,k}^{c}$, we have $TR_{m,n}^{f}=\max_{ k\in pre(n)}TF_{m,k}^{f}$, hence the objective function $F(P_{m,n,p}^{tx})$ can be rewritten as:
\begin{equation}\label{eq:case2}
\begin{split}
F(P_{m,n,p}^{tx})=&  \sum_{m=1}^{M} \sum_{n=1}^{N} \left\{TF_{m,n}^{f}+\mu_{m,n}\left[\tau_{m,n}^{t}+\max_{ k\in pre(n)}(TF_{m,k}^{l}-  \right. \right. \\
& \left. \left. TF_{m,k}^{f})\right] \right\}
\end{split}
\end{equation}

\textbf{Case III:} If $\max_{ k\in pre(n)}TF_{m,k}^{c}>TF_{m,n}^{t}$ and $\max_{k\in pre(n)} TF_{m,k}^{c}>\max_{ k\in pre(n)}TF_{m,k}^{f}$, we have $TR_{m,n}^{f}=\max_{k\in pre(n)}TF_{m,k}^{c}$. The objective function $F(P_{m,n,p}^{tx})$ can be rewritten as:
\begin{equation}\label{eq:case3}
\begin{split}
F(P_{m,n,p}^{tx})=& \sum_{m=1}^{M} \sum_{n=1}^{N}
\left\{TF_{m,n}^{f}+\mu_{m,n}\left[\tau_{m,n}^{t}+\max_{ k\in pre(n)}(TF_{m,k}^{l}- \right. \right. \\
&\left. \left. TF_{m,k}^{c})\right] \right\}.
\end{split}
\end{equation}
It is easy to verify that $F(P_{m,n,p}^{tx})$ is nonconvex w.r.t. $P_{m,n,p}^{tx}$, however, the optimal transmission power can be determined by adopting the maximum transmission power $P_{m,n}^{tx,max}$, which is the same for case II and case III of (\ref{eq:opa1}). In (\ref{eq:opa2}), there are also three cases as follows.\

\textbf{Case I':} If $TF_{m,n}^{t}+\tau_{m,n}^{r}>\max_{k\in pre(n)}TF_{m,k}^{c}$ and $TF_{m,n}^{t}+\tau_{m,n}^{r}>TF_{m,n}^{r}$, we have $TR_{m,n}^{c}=TF_{m,n}^{t}+\tau_{m,n}^{r}$. The objective function of (\ref{eq:opa2}) can be rewritten as:
\begin{equation}
\begin{split}
F(P_{m,n,p}^{tx})= &\sum_{m=1}^{M} \sum_{n=1}^{N} \left[TF_{m,n}^{c}+\lambda_{m,n}(TF_{m,n}^{r}-TF_{m,n}^{t}- \right. \\
&\left. \tau_{m,n}^{r})\right],
\end{split}
\end{equation}
where the minimum value can be achieved when the transmission power is $P_{m,n}^{tx,max}$.

\textbf{Case II':} If $\max_{k\in pre(n)}TF_{m,k}^{c}>TF_{m,n}^{t}+\tau_{m,n}^{r}$ and $\max_{k\in pre(n)}TF_{m,k}^{c}>TF_{m,n}^{r}$, we have $TR_{m,n}^{c}=\max_{k\in pre(n)}TF_{m,k}^{c}$. The objective function of (\ref{eq:opa2}) can be rewritten as:
\begin{equation}
\begin{split}
&F(P_{m,n,p}^{tx})=  \sum_{m=1}^{M} \sum_{n=1}^{N}
\left[TF_{m,n}^{c}+\omega_{m,n}(TF_{m,n}^{t}+\tau_{m,n}^{r}- \right.  \\
& \left.  \max_{k\in pre(n)}TF_{m,n}^{c})+\lambda_{m,n}(TF_{m,n}^{r}-\max_{k\in pre(n)}TF_{m,n}^{c})  \right],
\end{split}
\end{equation}
where the $F(P_{m,n,p}^{tx})$ is minimal when the transmission power is also $P_{m,n}^{tx,max}$.

\textbf{Case III':} If $TF_{m,n}^{r}>TF_{m,n}^{t}+\tau_{m,n}^{r}$ and $TF_{m,n}^{r}>\max_{k\in pre(n)}TF_{m,k}^{c}$, we have $TR_{m,n}^{c}=TF_{m,n}^{r}$. The objective function of (\ref{eq:opa2}) can be rewritten as:
\begin{equation}
\begin{split}
&F(P_{m,n,p}^{tx})=  \sum_{m=1}^{M} \sum_{n=1}^{N}
\left[TF_{m,n}^{c}+\omega_{m,n}(TF_{m,n}^{t}+\tau_{m,n}^{r}- \right.  \\
& \left.  \max_{k\in pre(n)}TF_{m,n}^{r})  \right],
\end{split}
\end{equation}
and similarly the device should transmit with its maximum power.

%% file: algoimp.tex
\subsection{Greedy Implementation}

\begin{algorithm}
\begin{algorithmic}[1]\caption{Greedy algorithm for mobile device $m$}\label{algo:greedy}
\REQUIRE{$\mathcal{N}$: a sequence of $N$ sub-tasks of mobile device $m$;\\
	                     $pre{(n)}$: the predecessors of task $n$;\\
	                     $\epsilon_m$: the sum budget for mobile device $m$;\\
	                     $P_{p}^{f}$: the charging price by the $p$th fog node;\\
	                      $P_{q}^{c}$: the charging price of the $q$th remote cloud server.\\}
		\ENSURE{Offloading policy $\mathbb{X}$}
            \STATE{\bf begin} \\
            \STATE{ Initialize: $w_{m,n}, d_{m,n}$, $U_{p}^{f}=0$; }\\
            \FOR{$n$ =$1$ to $N$}
              \STATE{Compute $R_{m,n,p}, \tau_{m,n}^{l}, E_{m,n}^{l}$ by (1)-(3), respectively; }\\
              \IF{$pre{(n)}==\varnothing$}
                 \STATE{$TR_{m,n}^{l}=0, TR_{m,n}^{f}=0, TR_{m,n}^{c}=0$}\\
                 \ELSE
                    \STATE{Compute $TF_{m,n}^{t}, TF_{m,n}^{r}$ by (17), (20), respectively;}\\
                    \STATE{Compute $TR_{m,n}^{f}, TR_{m,n}^{l}, TR_{m,n}^{c}$ by (12), (15), (16), respectively;}\\
                \ENDIF
                    \STATE{Compute $TF_{m,n}^{l}, TF_{m,n}^{f}, TF_{m,n}^{c}$ by (14), (18), (19), respectively;}\\
                    \IF{$TF_{m,n}^{l}< TF_{m,n}^{f}$ \& $TF_{m,n}^{l}< TF_{m,n}^{c}$}
                     \STATE{$x_{m,k,p}^{f}=0, x_{m,k,q}^{c}=0$;}\\
                     \ELSE
                         \IF{$\frac{P_q^{c}d_{m,n}}{E_{m,n}^{c}} \ge 1$}
                             \STATE{$x_{m,k,p}^{f}=0, x_{m,k,q}^{c}=1$;}\\
                          \ELSE
                               \STATE{$x_{m,k,p}^{f}=1, x_{m,k,q}^{c}=0$;}\\
                          \ENDIF
                     \ENDIF
                     \STATE{Compute $E_{m,n}^{f}, E_{m,n}^{c}$ by (5), (6), respectively;}\\
		             \STATE{Compute $C_{m,n}$ by (26);} \\
            \ENDFOR
            \WHILE{$\sum_{n=1}^{N} C_{m,n}> \epsilon_m $}
               \IF{$\sum_{n=1}^{N} x_{m,n,q}^{c} \neq 0$}
                   \STATE{$y== \underset{n}{arg} min\{E_{m,n}^{c}\}$;}\\
                   \STATE{$x_{m,y,p}^{f}=1, x_{m,y,q}^{c}=0$;}\\
               \ELSE
                   \STATE{ $z== \underset{n}{arg}  min\{E_{m,n}^{f}\}$; }\\
                   \STATE{$x_{m,z,p}^{f}=0, x_{m,z,q}^{c}=0$;}\\
               \ENDIF
            \ENDWHILE
            \WHILE{$ U_{p}^{f} < 0 $}
  	\IF{$\sum_{n=1}^{N} x_{m,n,q}^{c} \neq 0 \& \{ \frac{E_{m,n}^{s}}{E_{m,n}^{f}}>1 \} \neq \varnothing $}
  	 \STATE{$u== \underset{n}{arg} max\{ \frac{E_{m,n}^{s}}{E_{m,n}^{f}} \}$;} \\
  	   \STATE{$x_{m,u,p}^{f}=1, x_{m,u,q}^{c}=0$;}\\
  	\ELSE
      \STATE{   $v== \underset{n}{arg}  min\{ \frac{P_{p}^{f} d_{m,n}}{E_{m,n}^{f}}$\}}; \\
  	 \STATE{$x_{m,v,p}^{f}=0, x_{m,v,q}^{c}=0$;}\\
    \ENDIF
      \STATE{Compute $U_{p}^{f}$ by (23);}\\
  \ENDWHILE
	\STATE{\bf end}
  \end{algorithmic}
\end{algorithm}
Based on the above analysis, first of all, we design a greedy offloading policy to minimize the task completion time. To acquire the minimum finish time of all subtasks in the application on mobile device $m$, the minimum completion time of subtask $n$ is selected from $TF_{m,n}^{l}$, $TF_{m,n}^{f}$, and $TF_{m,n}^{c}$. To meet the utility constraint of the remote cloud server, $P_q^{c}d_{m,n}$ should be larger than $E_{m,n}^{c}$ according to (\ref{eq:enermncloud}). By determine offloading policy for each task iteratively,  we can get the initial offloading solution for mobile device $m$. This sub-procedure is shown between Line $3$ to Line $21$.
 	
 Then, we iteratively adjust the initial offloading policy to satisfy the sum cost budget for mobile device $m$. Considering the fact that in real scenarios, the price of remote cloud service, \ie $P_q^{c}$ is much more expensive than the fog node service price $P_{p}^{f}$. Therefore, the cost of a task on the cloud is higher than that on the fog node, and the cost of a task on the fog node is higher than that on the mobile device with respect to (\ref{eq:mdcost}). If there have been some tasks deployed on the cloud, \ie $\sum_{n=1}^{N}x_{m,n,q}^{c} \ne 0$, we can move them to the fog node one by one with respect to the ascending order of $E_{m,n}^{c}$. Similarly, when there is no task deployed on the cloud, we can move tasks from the fog node to the mobile device one by one in ascending order of $E_{m,n}^{f}$.

 Finally, we adjust the offloading policy to satisfy the utility constraint of the fog node. In order to improve the utility of the fog node, we remove tasks from cloud server to fog node in descending order of $\frac{E_{m,n}^{s}}{E_{m,n}^{f}}$, when $E_{m,n}^{s}$ is greater than $E_{m,n}^{f}$. If no task is deployed on cloud server, we remove tasks from fog nodes to mobile device in ascending order of $\frac{P_{p}^{f} d_{m,n}}{E_{m,n}^{f}}$. It should be noted that, with the above operations, the total cost budget for mobile device $m$ still holds, because of the decrease of ask price from remote cloud to fog and from fog to mobile device changes. The detail of the greedy algorithm is described in Algorithm \ref{algo:greedy}.

\begin{theorem}
For a particular application, the greedy algorithm on computation offloading takes time of $O(N)$.
\begin{proof}
The generation of initial offloading policy step takes time of $O(N)$, to meet the total cost budget, the adjusting step runs no more than $O(N)$, and it takes no more than $O(N)$ time to satisfy the utility constraint of the fog node. Therefore, the total time complexity of the greedy algorithm is $O(N)$, which concludes the proof.
\end{proof}
\end{theorem}

\subsection{Simulated Annealing Algorithm}
 \begin{algorithm}[h]
    \begin{algorithmic}[1]\caption{{(Simulated annealing algorithm for mobile device $m$)}}\label{algo:sa}
      \REQUIRE{$\mathcal{N}$: a sequence of $N$ sub-tasks of mobile device $m$;\\
  			$\epsilon_m$: the sum budget for mobile device $m$.}\\
       \ENSURE{Offloading policy $\mathbb{X}$}\\
       \STATE{\bf begin}\\
       \STATE{    Initialize: a random policy $\mathbb{X}$, $Tem=T0$, $cool= 0.98$, $U_{p}^{f}=0$, $U_{q}^{c}=0$.} \\
		 \WHILE{$Tem> 0.1$ \& $U_{p}^{f} \ge 0$ \& $U_{q}^{c} \ge 0$}
		   \STATE{Compute($C(\mathbb{X}), T(\mathbb{X})$);}\\
		 	\STATE{$\mathbb{X}_c \leftarrow \mathbb{X}$;}\\
		 	\STATE{$x\_change==random[-3,3]$;}\\
		     \STATE{$index==random[1,n]$;}\\
		     \STATE{${x\_c}_{index} == x_{index} + x\_change$;}\\
		       \IF{${x\_c}_{index}<1$}
		         \STATE{${x\_c}_{index}==1$;}\\
                \ELSE
                   \IF{${x\_c}_{index}>3$}
		            \STATE{${x\_c}_{index}==3$;}\\
                     \ENDIF
                \ENDIF
		         \STATE{$Tem==Tem*cool$;}\\
		        \STATE{ Compute($C(\mathbb{X}_c), T(\mathbb{X}_c)$);}\\
		   \IF{ accepted with probility $exp(\frac{-(T(\mathbb{X}_c)-T(\mathbb{X}))}{Tem})$}
               \STATE{ $\mathbb{X}\leftarrow \mathbb{X}_c$; }\\
               \STATE{ Compute $U_{p}^{f}, U_{q}^{c} $ by (23), (24), respectively;}\\
           \ENDIF
	     \ENDWHILE
        \WHILE{$C(\mathbb{X}) > \epsilon_m$}
        \STATE{Back to step 1;   }\\
        \ENDWHILE
  \end{algorithmic}
 \end{algorithm}
The simulated annealing technique (SA) was initially proposed to solve the hard combinatorial optimization problems.
It is a technique to find a better solution for an optimization problem by trying random variations of the current solution \cite{Fan2014Simulated}.

In the basic simulated annealing algorithm, an initial solution is always selected from the range of the parameters in random. We obtain the initial solution by randomly choose a list of number from $\{1,2,3\}$, where the number $1$ denotes that the task is executed on the mobile device, the number $2$ denotes that the task is offloaded to the fog node, and $3$ means the task is scheduled on the cloud server. A feasible neighboring solution is generated by randomly choose a task, and adjust the offloading policy to a different one. The simulated annealing algorithm consists of seven main steps, which is listed as follows.

Step $1$: Initialization (line $2$). We will design a set of parameters to control the process of SA, including the initial temperature $T0$, the final temperature $Tf$, the way of temperature reduction $\alpha$,  increment counter $L$ of a certain temperature, a randomly generated initial solution $\mathbb{X}$, and the utility of cloud server and fog nodes. Step $2$: Compute the total cost and the completion time of application in the current solution, \ie $C(\mathbb{X}), T(\mathbb{X})$. The detail is shown in line $5$. Step $3$: It generates a feasible neighboring solution $\mathbb{X}_c$, computes the sum cost of application and the finish time of application of new solution \ie $C(\mathbb{X}_c), T(\mathbb{X}_c)$. (lines $6$-$13$). Step $4$: Determine the Metropolis condition. If $T(\mathbb{X}_c)-T(\mathbb{X}) <0$, the solution is accepted. Otherwise, the solution is accepted with a probability of $exp(\frac{-(T(\mathbb{X}_c)-T(\mathbb{X}))}{Tem})$. Step $5$: Under the current temperature, the algorithm then checks the increment counter. If the iteration number is not the maximum one, then the algorithm will return to step $3$ as shown in line $6$. Step $6$: Reduce the temperature $Tem$. If the new temperature is greater than the stopping temperature $Tf$ and the utility constraints of both cloud server and fog node are met, then continue. Otherwise, the new solution is acquired. Step $7$: For the total cost of the new solution, the algorithm compares it with the budget. If the sum of cost cannot satisfy the budget, return to step $2$. Otherwise, the solution is accepted. The detail of the simulated annealing algorithm is described in Algorithm \ref{algo:sa}.

\subsection{Brute Force Method}
Since this is the first study on the optimal computation offloading with service cost and user budget considerations, there is no research to directly compare with. For comparing purpose, we implement a Brute Force computation offloading scheme. The Brute Force method explores all cases of offloading decisions and saves the one with the minimum complete time of the whole application. 

%% file: simulation.tex
\subsection{Simulation Setup}
To study the proposed algorithms, we design experimental studies on the offloading performances made by the corresponding algorithms. The results depicted in the figures are averaged for $1000$ time executions. We implement all the algorithms on the Matlab simulator. By default, we set the signal noise between the fog and mobile as $\sigma^2=1$, the wireless bandwidth is set as $W=5$MHz. The transmission bandwidth between fog node and remote cloud is set as $\omega=100$Kb. The CPU frequency of the fog node is set as $f_p^{f}=3.6\times10^{9}$Hz. The CPU frequency of the remote cloud node is set as $f_q^{f}=36\times10^{9}$Hz and the CPU frequency of the mobile device is $f_m^{l}=1\times 10^{9}$Hz. The amount of additional power is $P_0=0.1$W. The charging price of fog node is $P_{p}^{f}=0.001$ and the charging price of the remote cloud is $P_{q}^{c}=0.004$. System parameters $\alpha_f=0.5$, $\beta_f=0.4$, $\alpha_c=0.6$, $\alpha_f=0.6$. In all the simulations, the topology is a sequential graph with task dependency. However, it is easy to show that the designed algorithms are suitable for any directed acyclic graphs.
\subsection{Performance Analysis}
\begin{figure*}[htb]
    \centering
    \setlength{\abovecaptionskip}{-0.2cm}
    \setlength{\belowcaptionskip}{-1em}
    \subfigure[]{
        \label{fig:rtvsds}
        \includegraphics[width=0.2\textwidth]{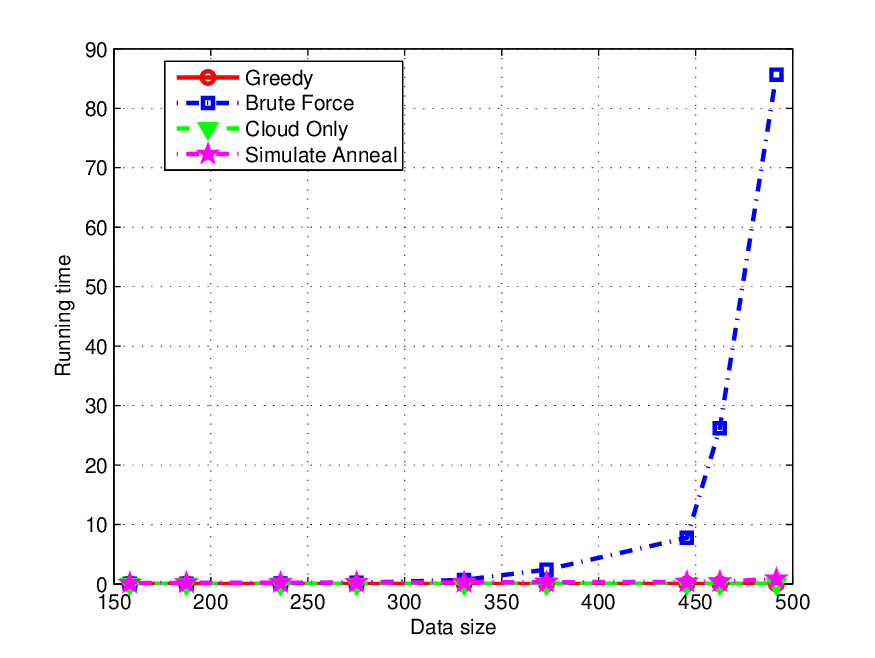}
    }
    \hspace{1mm}
    \subfigure[]{
        \label{fig:objvsds}
        \includegraphics[width=0.2\textwidth]{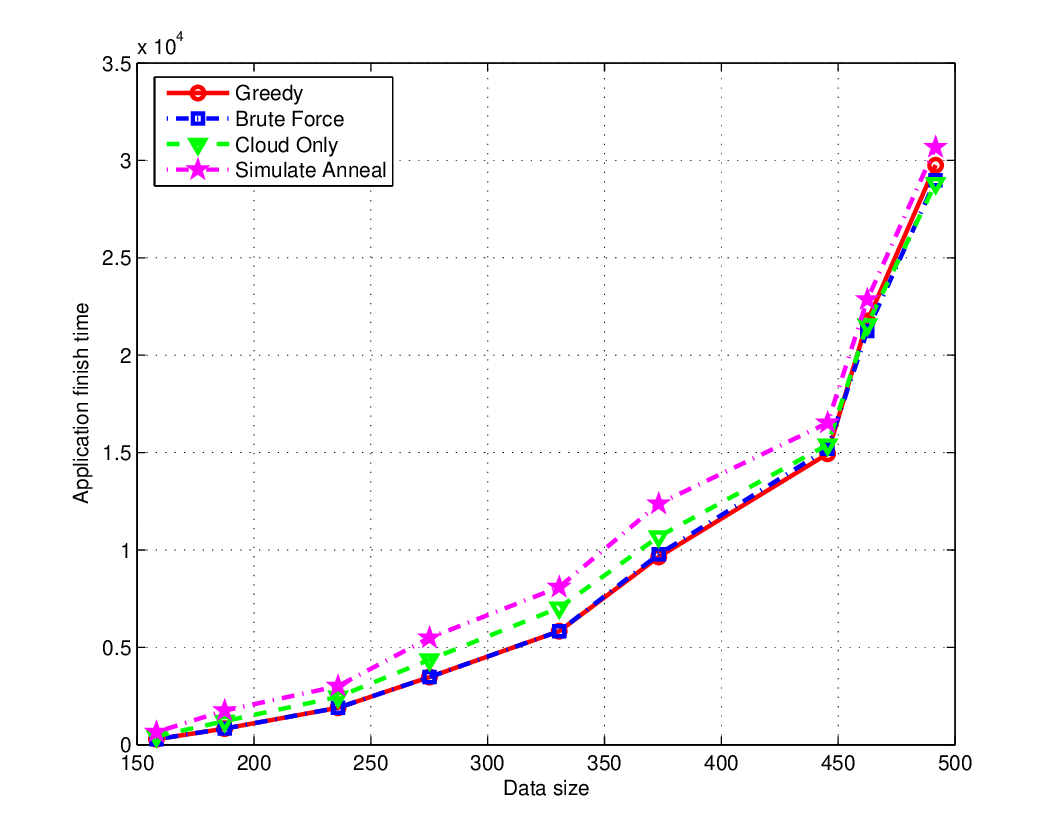}
    }
    \hspace{1mm}
    \subfigure[]{
        \label{fig:utilityUcUfvsds}
        \includegraphics[width=0.2\textwidth]{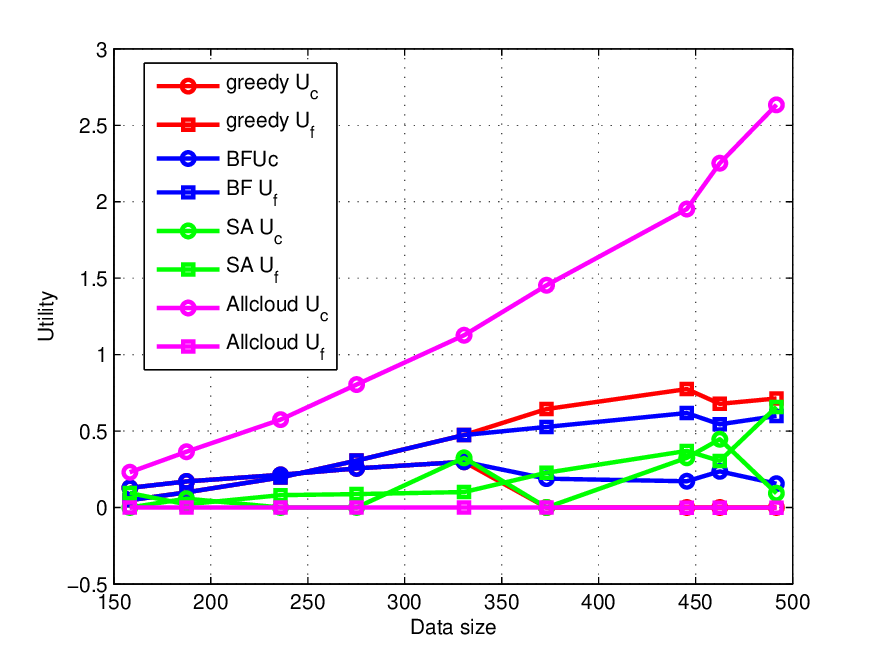}
    }
    \hspace{1mm}
    \subfigure[]{
         \label{fig:nooftaskvsbgt}
         \includegraphics[width=0.2\textwidth]{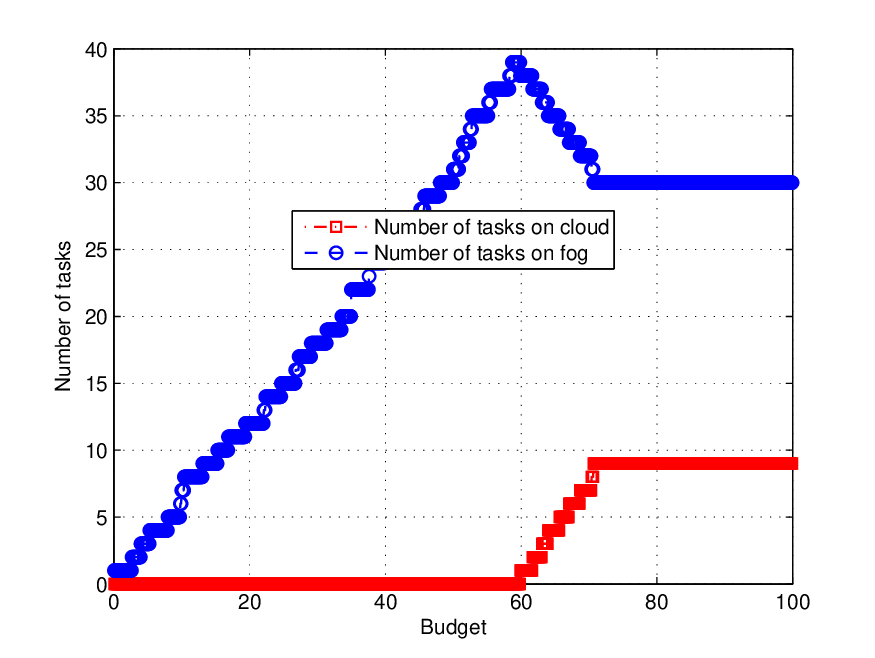}
    }
    \small \caption{(a) Running time Vs. data size; (b) The task finish time with the increasing of data size w/o budget; (c) Average utility of cloud $U_c$ and fog $U_f$ with data size; (d) Number of tasks with budget.}
\end{figure*}

The first experiment examines the total running time of all algorithms. As shown in Fig. \ref{fig:rtvsds}, with the increasing of data size, the running time of Brute Force method grows exponentially while the other three algorithms show linear running time growth. Then we study the relationship between application finish time and the workload. As shown in Fig. \ref{fig:objvsds}, the Brute Force method as well as the greedy algorithm demonstrate the best performance on the application finish time. The simulated annealing algorithm shows the highest application finish time. That's because the simulated annealing algorithm is a random probability based algorithm. The search direction may not always be the optimal one. On average, the Brute Force method and the greedy algorithm outperform the simulated annealing algorithm by about $28.13\%$. For the cloud only algorithm, the performance is meaningless, since it violates the constraints mentioned in the ENGINE problem. The third experiment examines the average utilities of both fog node and the remote cloud server, with the increasing of workload. Because that the constraints in $C4$ of (\ref{eq:obj}), for the all cloud case, $C4$ cannot be met, thus the values can be the upper bound of $U_p^{f}$ and the lower bound of $U_q^{c}$. With the increasing of workload, for the all cloud case, when all tasks are executed on the cloud server, the utility of cloud shows a linear growth. For the greedy algorithm, the utility of the cloud is close to zero. That's because for each subtask, the tight constraint of $U_q^{c}>0$ will motivate more subtasks to choose the fog. However, when $U_q^{c}>0$ cannot be met, the choice of moving some of subtasks from fog to local will result in the decrease of $U_p^{f}$. As shown in Fig. \ref{fig:nooftaskvsbgt},we set $N=40$. For the greedy algorithm, with increasing number of user budget, the number of allocated tasks on the fog increases first accordingly while the number of allocated tasks on the remote cloud stay constant at first. That's because the greedy scheme tries to utilize the fog node first. However, when the budget is higher than $59.6$, then the fog has reached its budget limit, thus it has to move some of the subtasks to the remote cloud, which results in an increase of the total number of tasks on the cloud and a decrease of the total number of tasks on the fog. Finally, the number of tasks on both fog and cloud will converge.

\begin{figure*}[htb]
    \centering
    \setlength{\abovecaptionskip}{-0.2cm}
    \setlength{\belowcaptionskip}{-1em}
    \subfigure[]{
        \label{fig:nooftasksvsapf}
        \includegraphics[width=0.2\textwidth]{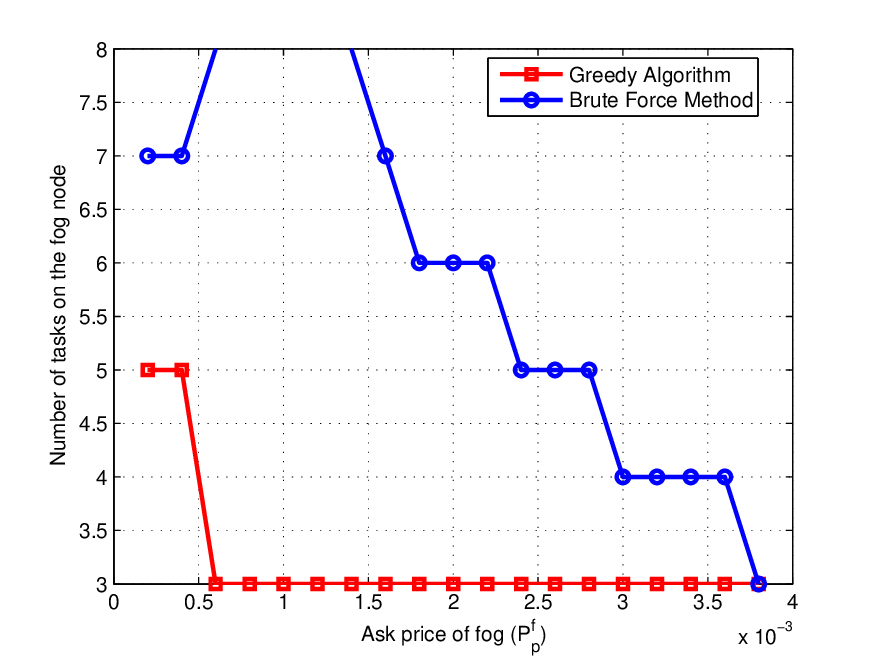}
    }
    \hspace{1mm}
    \subfigure[]{
        \label{fig:aftvsds}
        \includegraphics[width=0.2\textwidth]{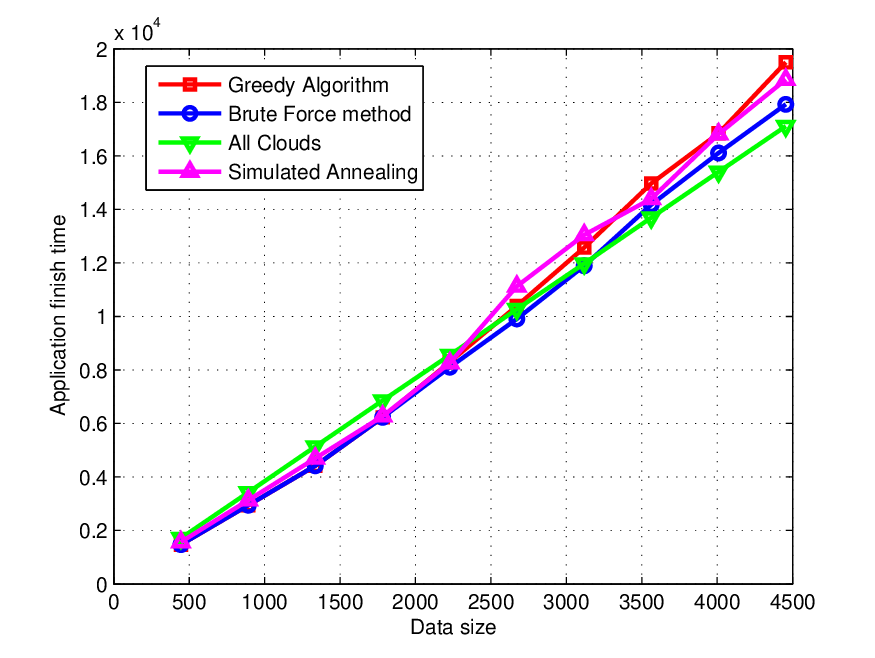}
    }
    \hspace{1mm}
    \subfigure[]{
        \label{fig:totalcostvsds}
        \includegraphics[width=0.2\textwidth]{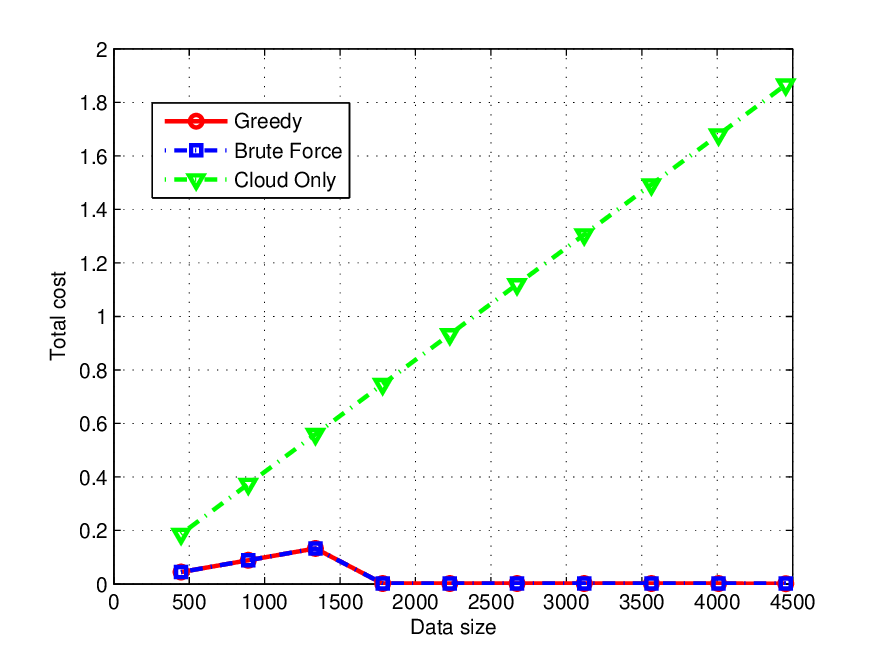}
    }

    \small \caption{(a) Number of tasks on the fog node with ask price of fog node; (b) The task finish time with the increasing of data size; (c) Total cost with data size.}
\end{figure*}

Next, we set user budget $\epsilon_m=6$. There are $9$ subtasks $[17.04, 87.60, 53.60, 29.19, 48.49, 39.20, 55.43, 42.56, 72.26]\times10$ required to be processed. As shown in Fig. \ref{fig:nooftasksvsapf}, when ask price of fog node increases, for both greedy algorithm and the Brute Force method, the number of tasks executed on the fog will decrease accordingly. When user budget is considered, as shown in Fig. \ref{fig:aftvsds}, the performance of all the proposed algorithms are close to the optimal all cloud scheme, which is different from Fig. \ref{fig:objvsds}. Finally Fig. \ref{fig:totalcostvsds} demonstrates that the greedy algorithm can save the cost and is close to the optimal Brute
Force method.


%% file: conclude.tex
This paper has addressed novel computation offloading schemes with user budget and service cost. The proposed models have shown their potential benefits to stimulate both service providers and mobile users.    